\documentclass[prl, aps, twocolumn, groupedaddress, superscriptaddress, nofootinbib]{revtex4-2}

\usepackage[dvipsnames]{xcolor}
\usepackage{graphicx}
\usepackage[normalem]{ulem} 
\usepackage{enumitem}       

\usepackage{amsmath}
\usepackage{amssymb}
\usepackage{bbm}            
\usepackage{newtxtext,newtxmath} 
\usepackage{bm}             

\usepackage[version=4]{mhchem} 
\usepackage{booktabs}
\usepackage{hhline}
\usepackage{array}
\usepackage{multirow}

\usepackage{tikz}

\usepackage[colorlinks=true,linkcolor=Blue,citecolor=Blue,urlcolor=Blue]{hyperref}

\newcommand\redout{\bgroup\markoverwith{\textcolor{red}{\rule[.5ex]{2pt}{0.4pt}}}\ULon}

\begin{document}
\title{Spin-orbit-induced Instability and Finite-Temperature Stabilization \linebreak of a Triangular-lattice Supersolid}

\author{Seongjun \surname{Park}}
\thanks{These authors contributed equally to this work.}
\affiliation{Department of Physics, Korea Advanced Institute of Science and Technology, Daejeon 34141, Korea}

\author{Sung-Min \surname{Park}}
\thanks{These authors contributed equally to this work.}
\affiliation{Department of Physics, Korea Advanced Institute of Science and Technology, Daejeon 34141, Korea}

\author{Yun-Tak \surname{Oh}}
\affiliation{Division of Semiconductor Physics, Korea University, Sejong 30019, Korea}

\author{Hyun-Yong \surname{Lee}}
\email{hyunyong@korea.ac.kr}
\affiliation{Division of Semiconductor Physics, Korea University, Sejong 30019, Korea}
\affiliation{Department of Applied Physics, Graduate School, Korea University, Sejong 30019, Korea}

\author{Eun-Gook \surname{Moon}}
\email{egmoon@kaist.ac.kr}
\affiliation{Department of Physics, Korea Advanced Institute of Science and Technology, Daejeon 34141, Korea}

\begin{abstract}
Geometrically frustrated triangular-lattice magnets provide fertile ground for realizing intriguing quantum phases such as spin supersolids. A common expectation is that spin–orbit coupling (SOC), which breaks continuous spin rotational symmetry, destabilizes these phases by gapping their low-energy modes. 
Revisiting this assumption, we map out the SOC–field phase diagram of a frustrated triangular-lattice magnet using spin-wave theory and infinite density-matrix renormalization group (iDMRG) simulations.
We find that while infinitesimally weak SOC indeed drives a zero-temperature instability of the supersolid by opening a gap, certain supersolid states remain thermodynamically stable at non-zero temperatures. This reveals a previously unrecognized mechanism in which thermal fluctuations counteract SOC to stabilize supersolidity. The resulting finite-temperature supersolids retain key responses, including a giant magnetocaloric effect, highlighting their potential relevance to real materials. At larger SOC, the system transitions into distinct magnetic orders, including a skyrmion lattice, completing a unified phase diagram.

\end{abstract}

\date{\today}
\maketitle

\section{Introduction}

Frustrated quantum magnets provide fertile ground for discovering unconventional quantum phases that emerge from the competition between exchange interactions and strong quantum fluctuations~\cite{sachdev2008, balents2010spin, starykh2015unusual, savary2016quantum}. Among these systems, triangular lattice antiferromagnets have served as prototypical platforms for realizing exotic many-body states, ranging from quantum spin liquids to spin supersolids~\cite{shimizu2003spin, Heidarian05, Wessel05, Melko05, zhou2012successive, zhou2017quantum, Tu22, gao2022spin, xiang2024giant}. The geometric frustration inherent to the triangular network suppresses conventional Néel order and produces an extensive manifold of nearly degenerate states, enabling subtle interplay between classical constraints and quantum zero-point motion. This delicate balance has motivated extensive efforts—both experimental and theoretical—to determine how these phases evolve under external tuning parameters such as magnetic fields, anisotropy, and pressure~\cite{zhou2012successive, Yamamoto2014, starykh2015unusual, zhou2017quantum}.

Recent studies on easy-axis triangular lattice antiferromagnets, most notably the nearly ideal Co-based compound \(\mathrm{Na}_2\mathrm{BaCo}{(\mathrm{PO}_4)}_2\), have uncovered compelling evidence for spin-supersolid behavior: a state in which longitudinal spin-density modulation coexists with transverse $U(1)$ phase coherence~\cite{gao2022spin, xiang2024giant}.
This magnetic analogue of the supersolid phase envisioned for helium provides a rare realization of coexisting crystalline and superfluid-like orders in a solid-state system~\cite{Boninsegni12}. 
The identification of the spin-supersolid behavior in \(\mathrm{Na}_2\mathrm{BaCo}{(\mathrm{PO}_4)}_2\) has, in turn, renewed interest in understanding the broader conditions under which supersolidity can remain stable in real magnetic materials.
Prior to these recent experimental discoveries, extensive theoretical studies had indeed predicted the emergence of supersolid phases arising from frustrated interactions in various contexts, ranging from extended Bose-Hubbard models~\cite{Batrouni2000, Hebert2001, Dong2017, Tu2020} and dipolar boson systems~\cite{CapogrossoSansone2010, Ohgoe2011, Ohgoe2012, Zhang2015, Wu2020, Zhang2021} to spin-dimer or triangular lattice antiferromagnets~\cite{Picon2008, Yamamoto2013, Yamamoto2014, Li2017, Liao2018, Zhu2019, tu2022}.

A central unresolved issue concerns the role of spin–orbit coupling (SOC). It is now well established that SOC can drive a wide range of emergent phenomena~\cite{witczakkrempa2014}, ranging from topological phases such as topological insulators, Weyl semimetals, and Dirac semimetals~\cite{hasan2010, armitage2018} to more exotic states including Kitaev spin liquids and non-Fermi liquids~\cite{ takagi2019, matsuda2025, lee2020, moon2013}. In frustrated magnets, SOC can profoundly modify exchange interactions and generate a variety of topological spin textures, including skyrmion lattices and chiral spin liquids~\cite{kurumaji2019skyrmion,broholm2020quantum}.

In the specific case of \(\mathrm{Na}_2\mathrm{BaCo}{(\mathrm{PO}_4)}_2\), SOC has often been regarded as negligible~\cite{xiang2024giant}. 
Two arguments underpin this expectation: (i) the estimated magnitude of SOC in Co-based  triangular lattice antiferromagnets is extremely small, and (ii) the classical spin configuration hosts a pseudo-Goldstone mode, suggesting that SOC should not open an appreciable gap at the harmonic level.
Nevertheless, symmetry dictates that SOC generally breaks continuous spin-rotational symmetry and forbids the transverse $U(1)$ coherence that underlies the supersolid order. This raises a fundamental question: To what extent can supersolidity survive in the presence of symmetry-allowed SOC, and what new phases might emerge as SOC grows?

Previous theoretical works have begun exploring SOC-driven phases in frustrated magnets, revealing the possibility of vortex crystals~\cite{Kamiya14}, multiple-Q states~\cite{Leonov15}, and skyrmion lattices even in systems where Dzyaloshinskii–Moriya interactions are forbidden by inversion symmetry~\cite{Hayami16, Lin16}. Yet, the stability of the supersolid phase~\cite{Heidarian05, Wessel05, Melko05} under SOC—and the nature of possible phase transitions out of it—remains poorly understood.
In particular, the interplay between SOC, quantum fluctuations, and thermal effects has not been systematically examined, leaving open whether realistic materials with moderate SOC can still host supersolid phenomenology.

Here we address this question by constructing the phase diagram of a triangular-lattice spin model incorporating symmetry-allowed SOC and an external magnetic field. Through a combination of symmetry analysis and spin-wave theory, we find that even infinitesimally weak SOC destabilizes the spin-supersolid phase at zero temperature, introducing a gap into its nominally gapless excitation mode. Using infinite density-matrix renormalization group (iDMRG) simulations, we further show that the system undergoes a sequence of SOC-driven quantum phase transitions into distinct magnetically ordered states, including a skyrmion-lattice phase at sufficiently strong SOC.

Beyond the zero-temperature instabilities, a key result of our work is the discovery that supersolidity can be thermodynamically stabilized at non-zero temperature despite the presence of SOC—a mechanism not captured by previous analyses. This finite-temperature regime reveals a constructive role of thermal fluctuations in partially restoring the effective symmetry associated with supersolid coherence, offering a new perspective on the robustness of supersolidity in frustrated magnets. Our findings also suggest that hallmark supersolid responses—such as the giant magnetocaloric effect—can persist even when SOC is not negligible, broadening the practical avenues for exploiting supersolid functionality in real materials.

Taken together, our results establish a unified SOC–field phase diagram for triangular-lattice magnets and provide a theoretical framework for interpreting recent experiments on Co-based triangular lattice antiferromagnets.
More broadly, they demonstrate how symmetry-allowed anisotropic interactions, thermal fluctuations, and magnetic fields cooperate to stabilize or reshape coherent quantum phases, opening new possibilities for realizing spin–orbit-driven topological and supersolid states in frustrated quantum materials.

\begin{figure}
    \centering
\includegraphics[width=0.99\columnwidth]{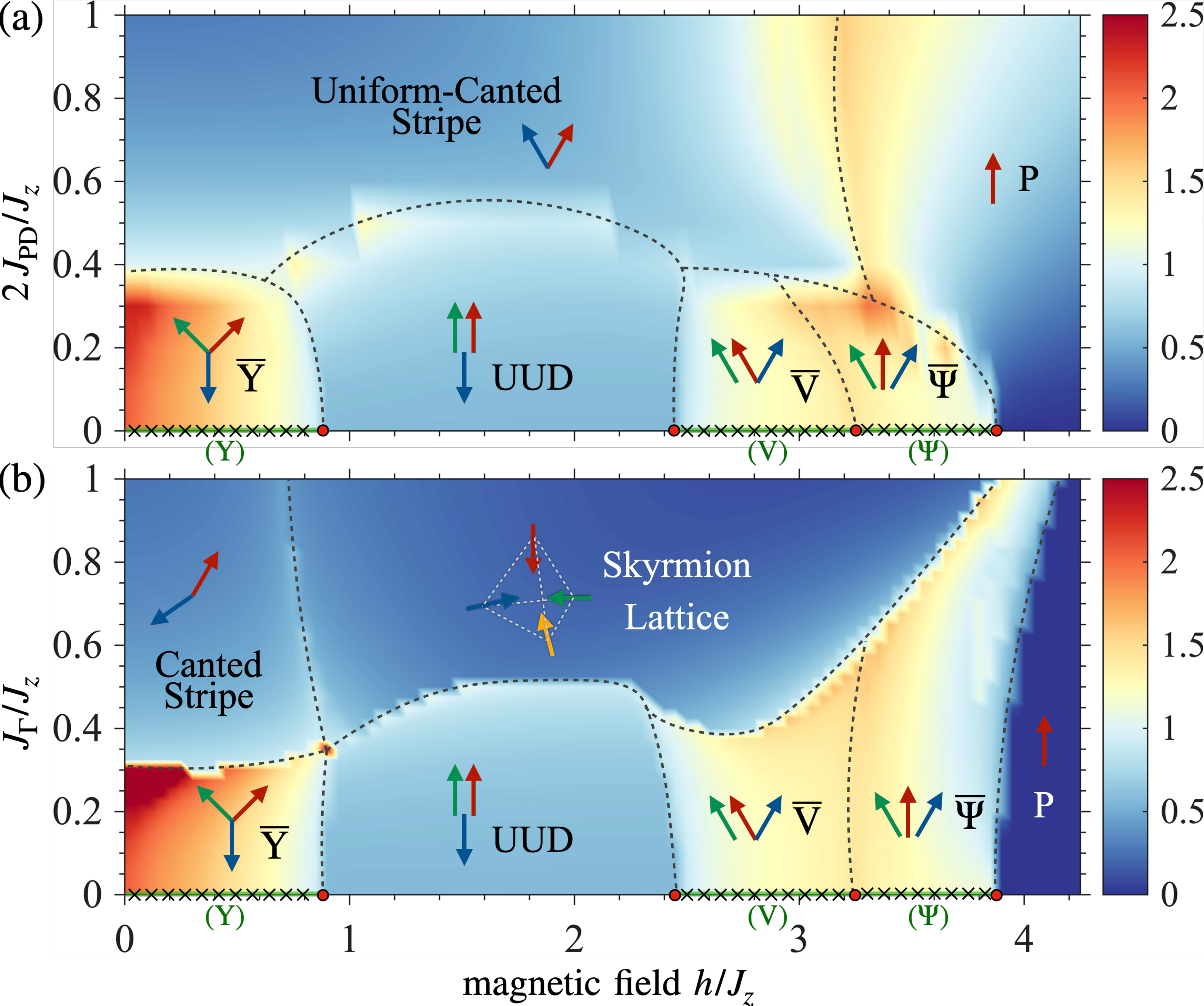}
    \caption{iDMRG phase diagrams of the triangular-lattice antiferromagnet with (a) the \(\Gamma\)-interaction and (b) the PD interaction at $J/J_z = 0.6$. Color encodes the maximal bipartite entanglement entropy of the ground state; dotted curves mark phase boundaries inferred from gradients of the energy density, magnetization, and entanglement entropy. The system is defined on an infinite cylinder with circumference \(W=6\). 
    The cross-hatched region corresponds to the spin-supersolid phase at zero temperature.
    }
    \label{fig:phase_diagram}
\end{figure}

\section{Results}

\paragraph{Model and symmetries.---}
We consider a spin-1/2 model on the two-dimensional triangular lattice~\cite{PhysRevLett.115.167203, PhysRevB.94.035107, PhysRevLett.120.207203, PhysRevX.9.021017, PhysRevB.96.054445, PhysRevLett.119.157201, 10.21468/SciPostPhys.4.1.003, PhysRevB.98.045119}, partly motivated by recent studies on \(\mathrm{Na}_2\mathrm{BaCo}{(\mathrm{PO}_4)}_2\) \cite{gao2022spin,xiang2024giant,PhysRevB.110.214408} in space group $P\overline{3}m1$, $\mathrm{A}_2\mathrm{Co}{(\mathrm{Se}\mathrm{O}_3)}_2$ ($\mathrm{A}$=$\mathrm{K}$ or $\mathrm{Rb}$) ~\cite{PhysRevMaterials.4.084406,chen2024phase, PhysRevLett.133.186704, PhysRevB.111.L060402,cui2025spin,zhu2025wannier} in $R\overline{3}m$, and \(\mathrm{YbMgGa}\mathrm{O}_4\) \cite{PhysRevX.9.021017,PhysRevB.94.035107,PhysRevLett.120.207203,PhysRevLett.119.157201,PhysRevLett.115.167203} in $R\overline{3}m$.
Given that the point group of the system is $\mathsf{{D}_{3d}}$, only four interactions ($J$, $J_z$, $J_{\rm PD}$, $J_{\rm \Gamma}$) are allowed in the presence of SOC by symmetry; detailed derivation is discussed in \cite{PhysRevX.9.021017,PhysRevLett.119.157201}. 

Our model Hamiltonian is
\begin{equation}\label{eq:1-model_Hamiltonian}
    H_{\rm tot} = H_{\rm XXZ} + H_{\rm PD} + H_{\Gamma},
\end{equation}
where  
\begin{eqnarray}
    H_{\rm XXZ} &=& \sum_{\langle ij \rangle} \left[ J \left( \hat{S}^{x}_{i} \hat{S}^{x}_{j} + \hat{S}^{y}_{i} \hat{S}^{y}_{j} \right) + J_z \hat{S}^{z}_{i} \hat{S}^{z}_{j} \right] - h \sum_i \hat{S}^{z}_{i}, \nonumber\\
    H_{\rm PD} &=& 2J_{\rm PD}  \sum_{\langle ij \rangle} \left( [\hat{S}^x, \hat{S}^y ]_{ij}\cos \varphi_{ij} - \lbrace \hat{S}^x, \hat{S}^y \rbrace_{ij} \sin \varphi_{ij} \right), \nonumber\\
    H_{\Gamma} &=& J_{\Gamma} \sum_{\langle ij \rangle} \left( \lbrace \hat{S}^y, \hat{S}^z \rbrace_{ij} \cos \varphi_{ij} - \lbrace \hat{S}^x, \hat{S}^z \rbrace_{ij} \sin \varphi_{ij} \right). \nonumber
\end{eqnarray}
The two operators, 
\([\hat{S}^\alpha, \hat{S}^\beta ]_{ij} \equiv \hat{S}^{\alpha}_{i} \hat{S}^{\alpha}_{j} - \hat{S}^{\beta}_{i} \hat{S}^{\beta}_{j}\) and \(\lbrace \hat{S}^\alpha, \hat{S}^\beta \rbrace_{ij} \equiv \hat{S}^{\alpha}_{i} \hat{S}^{\beta}_{j} + \hat{S}^{\beta}_{i} \hat{S}^{\alpha}_{j}\), are introduced, and $\langle ij \rangle$ denotes nearest-neighbor pairs on a two-dimensional triangular lattice. 
The angle \(\varphi_{ij} = 0, \pm \frac{2\pi}{3}\) depends on the orientation of the bond \(\langle ij \rangle\). We note that the three Hamiltonian terms ($ H_{\rm XXZ}$, $ H_{\rm PD}$ and $ H_{\Gamma}$) can be distinguished by their symmetry properties.

Three remarks are as follows.
First, the Zeeman coupling term in $H_{\rm XXZ} $ with an external magnetic field perpendicular to the triangular lattice, $h$, reduces the point-group symmetry from $\mathsf{{D}_{3d}}$ to $\mathsf{{S}_{6}}$. 
Second,  $H_{\rm PD} $ and $H_{\Gamma} $ from SOC explicitly breaks the $U(1)$ symmetry of $H_{\rm XXZ}$ but respect  the $\mathsf{{S}_{6}}$ symmetry once lattice points and spins are rotated {\it together} as a consequence of SOC.
For example,  a sixfold rotation followed by a reflection on the $xy$ plane, $S_6 \in \mathsf{{S}_6}$ transforms a spin operator,
\begin{eqnarray}
S_6: \quad S_j^{x} \pm i S_j^{y}  \rightarrow  e^{\mp{}i \frac{2\pi}{3}}(S_{j'}^{x} \pm i S_{j'}^{y} ), \quad S_j^{z}  \rightarrow S_{j'}^{z}, 
\label{eq:s6_transform}
\end{eqnarray}
where the lattice point index $j$ is also rotated into $j'$. Below, we demonstrate that the $S_6$ transformation plays a key role to determine thermal stability of a supersolid phase.
Third, the least symmetric term, $H_{\rm \Gamma}$, dictates the symmetry of the total Hamiltonian. 
In particular, $H_{\rm PD}$ remains invariant under a $\pi$ rotation, i.e., $[H_{\rm PD},\exp({-i\pi\sum_j S_j^z})]=0$, while $H_{\Gamma}$ is not invariant.

\begin{figure}[t]
    \centering
    \includegraphics[width=0.98 \columnwidth]{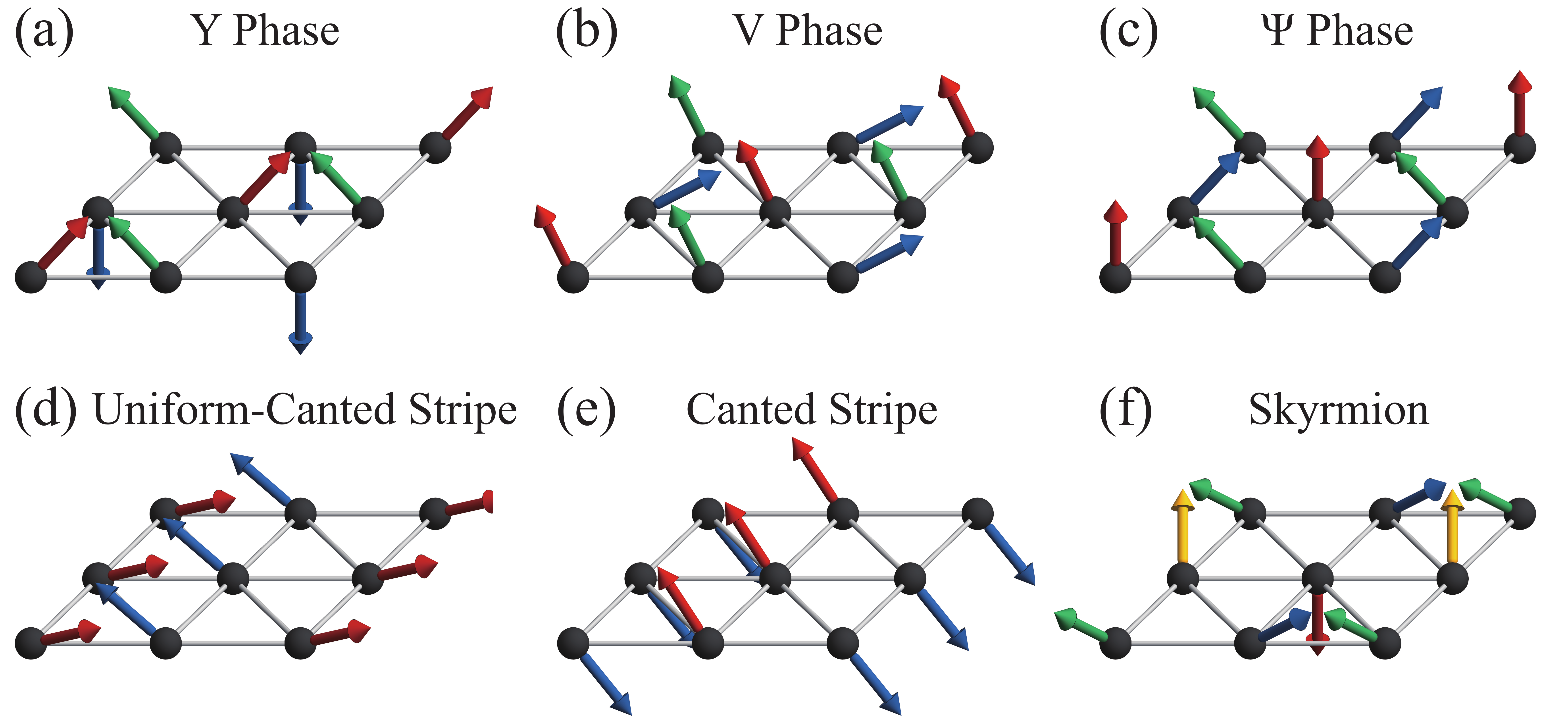}
    \caption{(a)-(f) Illustrations of magnetic orderings. 
    For states with three-sublattice order (a-c), we introduce sublattice labels (A, B, C) and use red, green, and blue to indicate the corresponding sublattices in the figures.}
    \label{fig:Spin_config}
\end{figure}
\paragraph{Ground state phase diagram.---}

To carve out the ground-state phase diagram of $H_{\rm tot}$, we employ the infinite density-matrix renormalization group (iDMRG)~\cite{White92,Schollwock05,Ian08, tenpy2024} on a Y-type cylinder with circumference $W=6$.

We examine two parameter-plane cuts of the model in Eq.~\eqref{eq:1-model_Hamiltonian}. Here, we set $J/J_{z} = 0.6$ which is relevant to \(\mathrm{Na}_2\mathrm{BaCo}{(\mathrm{PO}_4)}_2\)~\cite{sheng2022, chi2024dynamical, Sheng25}.
Figure~\ref{fig:phase_diagram}(a) shows the \((h,J_{\rm PD})\) plane at fixed \(J_\Gamma=0\), and Fig.~\ref{fig:phase_diagram}(b) shows the \((h,J_\Gamma)\) plane at fixed \(J_{\rm PD}=0\). In both panels, the background is a density map of the bipartite entanglement entropy evaluated at each parameter point, and the overlaid dashed lines mark phase boundaries inferred from the behavior of the energy density, magnetization, entanglement entropy, and their derivatives.

For cases without SOC-induced interactions ($J_{\rm PD} =J_{\rm \Gamma} =0$), our iDMRG calculations identify five distinct phases: three spin-supersolid phases (Y, V, and \(\Psi\)), a up-up-down \(1/3\)-magnetization plateau\,(UUD), and a polarized phase\,(P). 
Spin configurations in each phase are illustrated in Fig.\,\ref{fig:Spin_config}.
Alongside the iDMRG calculations, we apply linear spin wave theory (LSWT) \cite{PhysRev.58.1098,PhysRev.87.568,Toth_2015} to map out the ground-state phase diagram.
The most significant difference is that, while the iDMRG calculations confirm the existence of the $\Psi$ phase, LSWT does not identify the $\Psi$ phase as either a ground state or a metastable state.
We attribute its emergence in iDMRG to a quantum fluctuation not captured within LSWT.

The ground state manifold of each spin-supersolid phase can be parameterized by a single continuous variable $\phi$, which represents the angle of a global spin rotation about the $z$-axis. In the absence of SOC, the ground state energy is independent of this rotation, leading to a continuous degeneracy over the interval $\phi \in [0, 2\pi)$. Consequently, the ground state manifold of the spin-supersolid phase can be identified as a 1-sphere, namely $\mathcal{M}_{{\rm Y/V/}\Psi}\cong S^1$. For simplicity, we ignore the additional $Z_3$ degeneracy arising from the three-sublattice structure in this analysis.

We find that the three-sublattice orders (Y, V, and $\Psi$) remain robust against weak SOC perturbations, forming finite ground state phases as presented in Fig.\,\ref{fig:phase_diagram}. For clarity in the following discussion, we adopt the notation $\overline{\rm Y}$, $\overline{\rm V}$ and $\overline{\Psi}$ for those phases to differentiate them from their $J_{\rm PD}=J_{\rm \Gamma}=0$ counterparts. The necessity of this distinction will become evident as we discuss their stability analysis below.

\begin{figure}[t]
    \centering
    \includegraphics[width=0.98 \columnwidth]{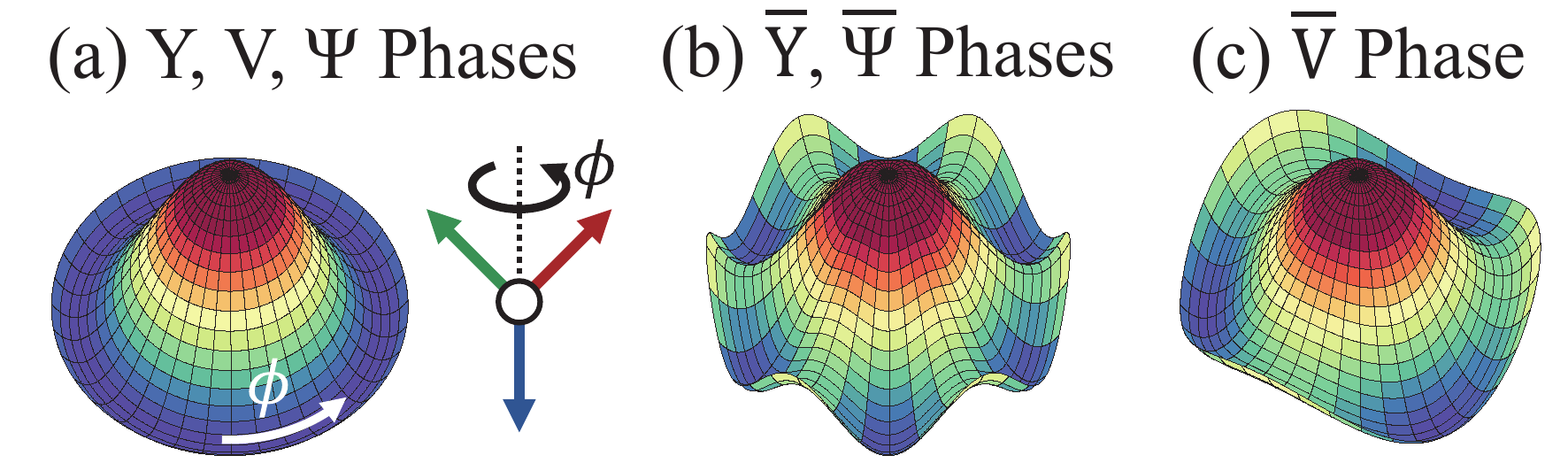}
    \caption{Illustration of the ground state manifolds. 
    (a)  In the absence of SOC, the ground state manifolds of three phases form a $S^1$-manifold, reflecting the continuous $U(1)$ degeneracy.
    In the presence of SOC, this manifold is distorted by discrete anisotropies: a six-fold ($Z_6$) for $\overline{\rm Y}$ and $\overline{\Psi}$ as in (b) and a three-fold ($Z_3$) for $\overline{\rm V}$  as in (c).} 
    \label{fig:GSM}
\end{figure}
As the magnitude of either $J_{\rm PD}$ or $J_\Gamma$ increases, the characteristic three-site sublattice orders becomes unstable and gives way to phases with modified magnetic unit cells. The associated spin textures are qualitatively reconstructed as well.

For finite $J_{\rm PD}$, the system develops a stripe order with a two-site magnetic unit cell, in which spins are ferromagnetically aligned along each stripe but adjacent stripes are antiferromagnetically ordered. At zero field $(h=0)$, all spins lie collinearly within the $xy$ plane. A magnetic field induces a uniform canting toward the field direction, generating a finite out-of-plane magnetization while maintaining the in-plane antiferromagnetic pattern as depicted in Fig.~\ref{fig:Spin_config}(d). We refer to this phase as the uniformly-canted stripe (UCS) phase. The canting grows smoothly with increasing $h$, continuously connecting the UCS to the P phase.

A finite $J_\Gamma$ similarly drives the emergence of a stripe-like order from the parent $\overline{\rm Y}$ phase, yet with a canting structure distinct from the UCS phase. At $h=0$, the spins are collinear in a plane tilted away from the $xy$ plane by the $\Gamma$ anisotropy [Fig.~\ref{fig:Spin_config}\,(e)]. Under the magnetic field, the spins cant non-uniformly, producing a modulated $z$-magnetization across the stripes; we refer to this state as the canted stripe (CS) phase.
We note that iDMRG calculations with $W=6$ cylinder initially find a non-magnetic ground state between the $\overline{\rm Y}$ and CS phases, consistent with the results reported in Ref.\,\onlinecite{PhysRevLett.120.207203}. However, this phase is destabilized on cylinders with larger circumferences, where the $\overline{\rm Y}$ phase extends to occupy the corresponding regime.

\begin{figure}[t]
    \centering
    \includegraphics[width=0.98 \columnwidth]{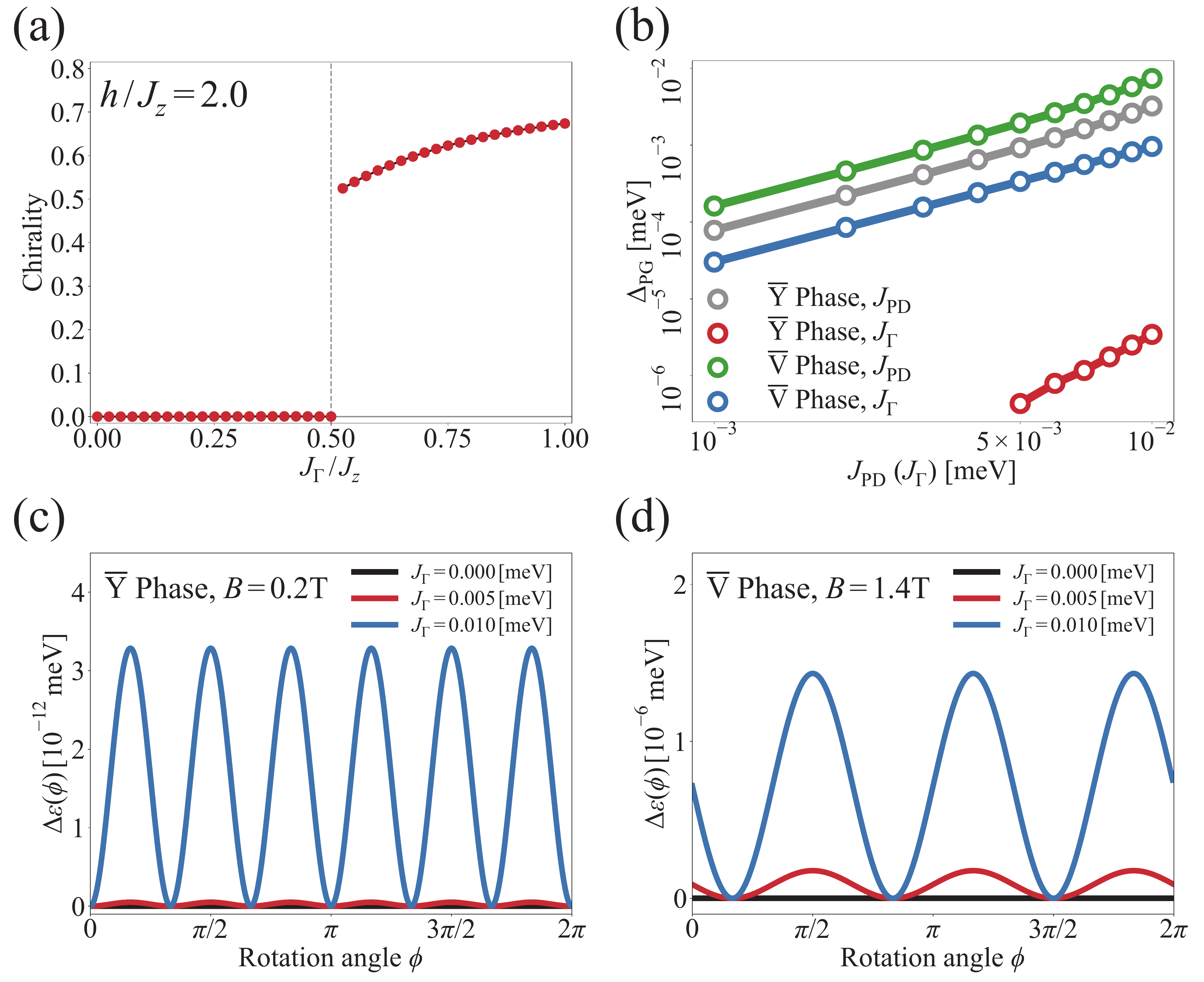}
    \caption{(a) Ground-state scalar spin chirality obtained from iDMRG as a function of \(J_{\Gamma}\) at fixed field \(h/J_{z}=2.0\) and $J_{\rm PD}=0$, showing the onset of chirality upon entering the SkX from the UUD state.
    (b) (gray and green) Pseudo-Goldstone gap \(\Delta_{\rm PG}\) as a function of \(J_{\rm PD}\) at \(J_{\rm \Gamma} = 0\). 
    (red and blue) \(\Delta_{\rm PG}\) as a function of \(J_{\rm \Gamma}\) at \(J_{\rm PD} = 0\).
    (c and d) Change in the semiclassical energy density \(\Delta\epsilon(\phi)=\epsilon(\phi)-\epsilon(\phi^{*})\) under a uniform spin rotation \(U_{z}(\phi)\), obtained from LSWT. 
    The angle $\phi^{*}$ denotes the position of local minima with respect to the $U(1)$ rotation.
    We fix \(J_{\rm PD}=0\), with varying \(J_{\Gamma}\) for the $\overline{\rm Y}$ phase (\(B = 0.2\,\rm T\)) and the $\overline{\rm V}$ phase (\(B = 1.4\,\rm T\)).
    We adopt the parameters $J = 0.075$~meV, $J_z = 0.125$~meV, and $h = g_z \mu_{\mathrm{B}} B$ with $g_z = 4.645$, appropriate for \(\mathrm{Na}_2\mathrm{BaCo}{(\mathrm{PO}_4)}_2\)~\cite{sheng2022, chi2024dynamical, Sheng25}
    }
    \label{fig:PG_Gap}
\end{figure}

In contrast, the $J_\Gamma$ interaction drives the UUD, V, and $\Psi$ phases into a quantum skyrmion-lattice (SkX) phase. The SkX exhibits an ultra-short magnetic period, with a four-site unit cell as illustrated in Fig.~\ref{fig:Spin_config}\,(f). Although its energetics are primarily classical, we find that the SkX is distinguished by a quantum mechanical multipolar operator, scalar spin chirality, $\chi = (1/N_\triangle)\sum_{\langle ijk\rangle \in \triangle}\big\langle \vec{S}_i \cdot (\vec{S}_j \times \vec{S}_k) \big\rangle$, where $\langle ijk\rangle$ runs over oriented nearest-neighbor triangles in the counterclockwise sense. Across the UUD--SkX transition, $\chi$ displays a finite jump and remains large throughout the SkX phase as presented in Fig.\,\ref{fig:PG_Gap}\,(a).

\paragraph{Instability of the spin-supersolid phases.---}
To demonstrate the instability of the spin-supersolid phases under SOC, we investigate their excitation spectra and ground state manifolds.

It is well understood that an accidental $U(1)$ symmetry nevertheless emerges in the spin-supersolid phases, even with $J_{\Gamma} \neq 0$ and $J_{\rm PD} \neq 0$, because the spin configurations in the spin-supersolid phases preserve three-fold rotational symmetry, leading to a cancellation of the anisotropic contributions from $H_{\rm PD}$ and $H_{\Gamma}$.
This is one well known example of a \emph{pseudo-Goldstone mode}~\cite{PhysRevLett.29.1698,PhysRevLett.121.237201}, leading to gapless excitations in the spin-wave spectrum within the non-interacting approximation. 
To overcome such artifacts, it was suggested to consider quantum fluctuations, manifestation of the \emph{order-by-quantum-disorder} mechanism~\cite{villain1980order,PhysRevLett.62.2056,shender1982antiferromagnetic} with the expression of the pseudo-Goldstone gap, $\Delta_{\rm PG} = (1/S) \sqrt{  (\partial_\theta^{2} \epsilon)_0  (\partial_\phi^{2} \epsilon)_0 -  (\partial_\theta \partial_\phi \epsilon)^2_0 }$, see Ref.~\cite{PhysRevLett.121.237201} for the precise expression.

Figure~\ref{fig:PG_Gap}(c) and (d) present the semiclassical energy density as a function of $\phi$ in the $\overline{\rm Y}$ and $\overline{\rm V}$ phases. 
They show that, for finite $J_{\rm \Gamma}$, quantum zero-point fluctuations lift the $U(1)$ degeneracy, selecting six and three specific directions as the ground states in the $\overline{\rm Y}$ and $\overline{\rm V}$ phases, respectively. 
Consequently, with finite $J_{\rm \Gamma}$ interaction, the ground state manifolds of the Y and V phases are reduced from $S^1$ to $\mathcal{M}_{\overline{\rm Y}} \cong Z_6$ and $\mathcal{M}_{\overline{\rm V}} \cong Z_3$, respectively.
Although Fig.~\ref{fig:PG_Gap}(c) and (d) are computed for $J_{\rm PD}=0$, the symmetry structure of the ground state manifolds remains unchanged for nonzero $J_{\rm PD}$, since the symmetry of the total Hamiltonian is dictated by its least symmetric term, $H_{\rm \Gamma}$.
The absence of the $\overline{\Psi}$ phase in the LSWT phase diagram precludes a direct LSWT analysis of the pseudo-Goldstone gap, we infer that the ground state manifold of $\overline{\Psi}$ phase becomes $Z_6$ in the presence of the SOC, analogous to the $\overline{\rm Y}$ phase, due to their shared symmetry characteristics. 
Using the semiclassical energy density, we calculate the pseudo-Goldstone gap, as shown in Fig.\,\ref{fig:PG_Gap}\,(b). The results demonstrate that the gap increases algebraically with the strength of the SOC perturbations.

Two remarks are particularly noteworthy.
First, the ground state manifolds of $\overline{\rm Y}$, $\overline{\rm V}$, and $\overline{\Psi}$ phases are not continuous, either $Z_6$ or $Z_3$ and thus $\overline{\rm Y}$, $\overline{\rm V}$, and $\overline{\Psi}$ phases are not supersolids any more at zero temperature. In other words, the presence of SOC naturally opens pseudo-Goldstone gap because SOC does not respect any continuous symmetries. 
Second, the difference between $\mathcal{M}_{\overline{\rm Y}}$ and $\mathcal{M}_{\overline{\rm V}}$ may be understood by the action of $S_6 \in \mathsf{{S_6}}$ transformation. 
Namely, the $S_6$ symmetry operation both rotates the lattice point and the spin direction as shown in Eq.\,\eqref{eq:s6_transform}.
When $S_6$ is applied to the C sublattice, the $\overline{\rm Y}$ phase configuration effectively rotates $\phi \rightarrow \phi +\pi/3$ while the $\overline{\rm V}$ phase configuration $\phi \rightarrow \phi -2\pi/3$.
The key point is that the sublattice points, A and B in Fig. 2(a), are effectively swapped in $\overline{\rm Y}$ phase while they are intact in $\overline{\rm V}$ phase in Fig. 2(b).
We note that the minimal angles of each spin supersolid phase are not universal but determined by the sign structures of the exchange interaction terms.

\paragraph{Finite-temperature stabilization of the spin-supersolid phases.---}
Computing the exact finite-temperature properties of the full Hamiltonian is generally intractable due to strong interactions and extensive degeneracies. Instead, we rely on universality, which indicates that systems sharing the same symmetry-breaking pattern and ground-state manifold exhibit identical long-wavelength thermal behavior. 

By constructing and analyzing alternative effective Hamiltonians within the same universality class, we capture the essential thermal physics of spin-supersolid phases while avoiding unnecessary microscopic complexity.
For example, the effective Hamiltonian describing the $\overline{\rm Y}$ phase may be written as
\begin{eqnarray}
\mathcal{H}_{\overline{\rm Y}}
= - \tilde{J} \sum_{\langle l,m\rangle}
\cos(\tilde{\phi}_l - \tilde{\phi}_m)
- \tilde{g}_6 \sum_{l} \cos(6\tilde{\phi}_l),
\end{eqnarray}
capturing $\mathcal{M}_{\overline{\rm Y}} = Z_6$.
Here, $\tilde{\phi}_l$ denotes coarse-grained fields defined on coarse-grained lattice sites, and the renormalized couplings $\tilde{J}$ and $\tilde{g}_6$ encode, respectively, the energy scale of the Y phase without SOC and the SOC-induced sixfold anisotropy.  
Although $\mathcal{H}_{\overline{\rm Y}}$ is purely classical in contrast to the full Hamiltonian $H_{\rm tot}$, the universality of thermal transitions guarantees the structure of the thermal transitions,~\cite{Sachdev_2011, Cardy_1996}.

The thermal behavior of $\mathcal{H}_{\overline{\rm Y}}$ has been extensively studied with various methods including the generalized Villain models~\cite{Jose_1978}.
Standard renormalization-group (RG) analysis in two spatial dimensions yields a Kosterlitz–Thouless (KT) phase for $T < \pi \tilde{J}/2$ when $\tilde{g}_6 = 0$.
Including the sixfold anisotropy leads to the RG flow
\begin{eqnarray}
\frac{d\tilde{g}_6}{dl}
= \left(2 - \frac{9T}{\pi \tilde{J}}\right)\tilde{g}_6,
\qquad
\tilde{g}_6(l)
= e^{\left(2 - \frac{9T}{\pi \tilde{J}}\right)l}\tilde{g}_6(0), \nonumber
\end{eqnarray}
where $l$ is the logarithmic length scale of the RG.  
Consequently, $\tilde{g}_6$ becomes irrelevant for $T > 2\pi \tilde{J}/9$, stabilizing a KT phase in the window
\[
\frac{2\pi\tilde{J}}{9} < T < \frac{\pi\tilde{J}}{2}.
\]
This thermally stable KT phase is fully consistent with our LSWT calculations: although $\tilde{g}_6$ is relevant at zero temperature, it becomes irrelevant at sufficiently high temperatures.  
The KT phase corresponds to the finite-temperature continuation of the Y phase, exhibiting quasi-long-range order characteristic of two-dimensional supersolids.  
Thus, the zero-temperature $\overline{\rm Y}$ phase evolves into a supersolid Y phase at finite temperatures and eventually into a paramagnetic state, as illustrated in Fig.~5(a).  
Since the $\Psi$ phase shares the same symmetry structure as the Y phase, its thermal phase diagram is expected to be analogous.

A similar construction applies to the $\overline{\rm V}$ phase, whose effective Hamiltonian is given by
\begin{eqnarray}
\mathcal{H}_{\overline{\rm V}}
= - \tilde{J}' \sum_{\langle l,m\rangle}
\cos(\tilde{\phi}_l - \tilde{\phi}_m)
- \tilde{g}_3 \sum_{l} \sin(3\tilde{\phi}_l),
\end{eqnarray}
which correctly yields $\mathcal{M}_{\overline{\rm V}} = Z_3$.
RG analysis shows that the threefold anisotropy eliminates any stable KT phase, thereby suppressing supersolidity altogether.  
Therefore, once spin-orbit coupling is present, the V phase cannot survive at finite temperatures, as illustrated in Fig.~5(b).

Note that the effective Hamiltonians employed here should be further scrutinized before being used for quantitative predictions in real materials. Nevertheless, the existence or absence of thermal spin-supersolid phases follows unambiguously from our analysis, as it is dictated by symmetry and universality rather than microscopic parameters.

\section{Discussion}

\begin{figure}
    \centering
    \includegraphics[width=0.99 \columnwidth]{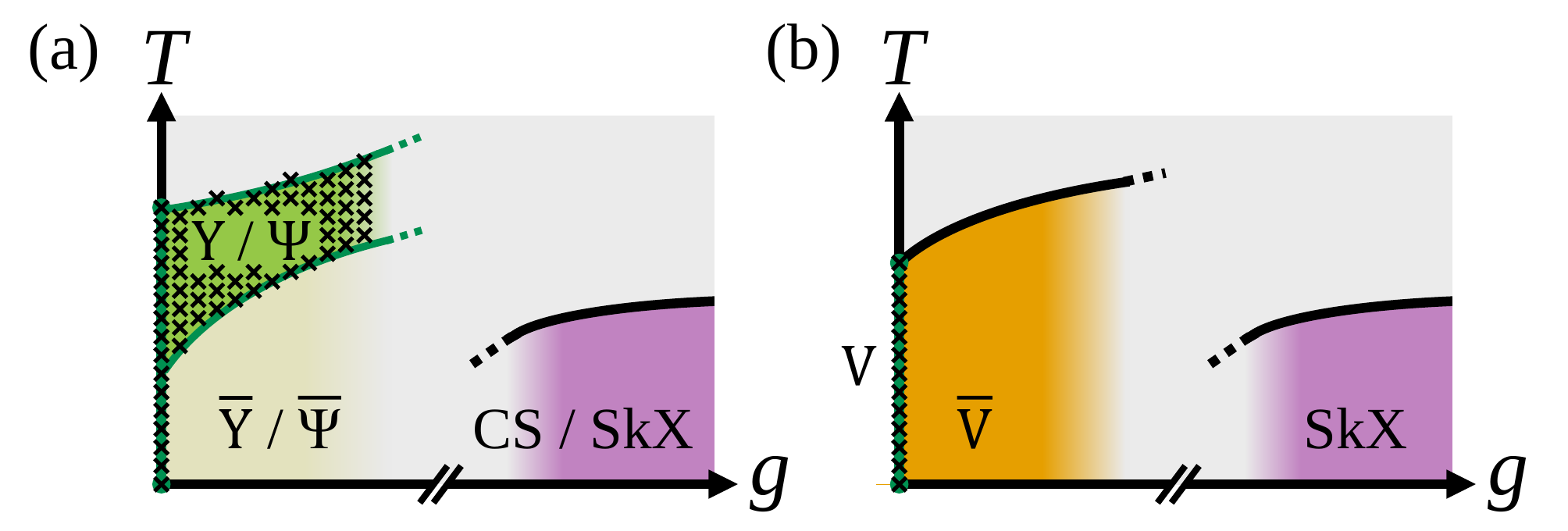}
        \caption{Schematic finite-temperature phase diagrams and the evolution of the ground state manifold. Phase diagrams for (a) $\overline{\rm Y}$ and $\overline{\Psi}$ phases and (b) the $\overline{\rm V}$ phase as a function of SOC strength $g$. The supersolid phases are indicated by the regions with $\times$. }

\label{fig:finite_T}
    \label{fig:1}
\end{figure}

In this work, we have comprehensively clarified the nature of symmetry-breaking phases in triangular lattice antiferromagnets and their stability against spin-orbit coupling by combining iDMRG simulations, LSWT, and renormalization group analysis.
Specifically, our LSWT calculations explicitly demonstrate the opening of a pseudo-Goldstone gap, confirming that the continuous $U(1)$ degeneracy is lifted in the presence of SOC.

Building on this, our RG analysis reveals a sharp dichotomy in the thermodynamic fate of the pseudo-Goldstone modes. For the Y and $\Psi$ phases, the SOC-induced $Z_6$ anisotropy becomes irrelevant at intermediate temperatures, allowing thermal fluctuations to restore quasi-long-range coherence by suppressing the discrete anisotropy. This leads to the emergence of a finite-temperature spin-supersolid phase bounded by a Berezinskii--Kosterlitz--Thouless transition. In contrast, the V phase experiences a relevant $Z_3$ anisotropy that prevents such restoration, eliminating any possibility of thermal supersolidity.

A particularly important implication of these results concerns the magnetocaloric effect, which has recently attracted attention due to the observation of a giant magnetocaloric effect in the spin-supersolid candidate \(\mathrm{Na}_2\mathrm{BaCo}{(\mathrm{PO}_4)}_2\)~\cite{xiang2024giant}.
In magnetic supersolid phases, the coexistence of soft transverse modes and longitudinal density modulations enhances magnetic entropy near phase boundaries, enabling large adiabatic temperature variations under field sweeps. A prevailing expectation is that SOC, by gapping these collective modes, should substantially suppress such entropy-driven responses.

Our results, however, reveal a more nuanced picture. Although SOC indeed gaps the pseudo-Goldstone modes at zero temperature, the thermal stabilization of the Y and $\Psi$ supersolids implies that the associated entropic enhancement remains active over a broad finite-temperature window. In this regime, the effective irrelevance of the six-fold anisotropy restores soft collective excitations characteristic of supersolidity, thereby preserving the conditions required for a large magnetocaloric effect. This mechanism provides a natural explanation for the persistence of the giant magnetocaloric response observed experimentally, even in systems where SOC is symmetry-allowed and not negligibly small. Moreover, the proximity to SOC-driven competing phases such as the uniformly canted stripe and skyrmion-lattice states can enhance entropy accumulation near multicritical regions, potentially amplifying magnetocaloric effect signatures.

Collectively, our results provide a unified framework, supported by both semiclassical calculations and iDMRG simulations, for understanding how explicit symmetry breaking, thermal fluctuations, and emergent discrete anisotropies shape the phase diagram of frustrated magnets with SOC. In particular, the SOC-magnetic field phase diagram constructed in this work offers a theoretical basis for interpreting the diverse magnetic responses of Co-based triangular-lattice antiferromagnets, including the persistence of supersolid-driven magnetocaloric effects. We hope that our findings stimulate further experimental exploration of entropy-based probes in systems where competing anisotropies and thermal restoration mechanisms intertwine to stabilize unconventional quantum phases.

\section{Acknowledgments}
We thank Jae Hoon Kim and Yuan Wan for insightful discussions.
H.-Y.L was supported by the Basic Science Research Program through the National Research Foundation of Korea funded by the Ministry of Science and ICT [Grant No. RS-2023-00220471, RS-2025-16064392].
S.P., S.-M.P., and E.-G.M. were supported by the National Research Foundation of Korea (NRF) grant funded by the Korea government (MSIT) (Grant No. RS-2025-00559286); 
by the Nano \& Material Technology Development Program through the NRF funded by MSIT (Grant Nos. RS-2024-00451261 and RS-2023-00281839); 
by the National Measurement Standards Services and Technical Support for Industries funded by the Korea Research Institute of Standards and Science (KRISS) (KRISS-2025-GP2025-0015);
and by the Global Partnership Program of Leading Universities in Quantum Science and Technology through the NRF funded by MSIT (Grant Nos. RS-2025-08542968 and RS-2023-00256050)

\bibliography{references}

\end{document}